\documentstyle[12pt]{article}

\begin{document}

\begin{titlepage}

\begin{flushright}
December 15, 1993
\end{flushright}

\vskip 36pt
\begin{center}
\Large\bf
ISING-LINK QUANTUM GRAVITY\\
\end{center}
\vskip 36pt

\begin{center}
  {\sl Tom Fleming \footnote[1] {Electronic address:
fleming@physics1.natsci.csulb.edu} and Mark Gross
\footnote[2] {Electronic Address: mgross@csulb.edu}\\
  Dept. of Physics and Astronomy,
  California State Univ.,
  Long Beach, CA 90840}
\vskip10pt

{\sl Ray Renken \footnote[3] {Electronic Address:
rlr@phys.physics.ucf.edu}\\
   Department of Physics,
   University of Central Florida,
   Orlando, Florida 32816}
\end{center}

\vfill

\begin{center}\bf Abstract \end{center}

We define a simplified version of Regge quantum gravity where
the link lengths can take on only 2 possible values, both always
compatible with the triangle inequalities.  This is therefore
equivalent to a model of Ising spins living on the links of a regular
lattice with somewhat complicated, yet local interactions.
The measure corresponds to the
natural sum over all $2^{\# \hskip2pt links}$ configurations,
and numerical simulations
can be efficiently implemented by means of look-up tables.  In
three dimensions we find a peak in the ``curvature susceptibility''
which grows with increasing system size.  However, the value of the
corresponding critical exponent as well as the behavior of the
curvature at the transition differ from
that found by Hamber and Williams for the Regge theory with
continuously
varying link lengths.

\vskip10pt

PACS number(s): 04.60.+n,11.15.Ha

\vfill

\end{titlepage}

{\bf I. INTRODUCTION}
\vskip15pt

To date, two main formulations of lattice quantum gravity have
been considered, the so called ``Regge gravity'' \cite{1,2}
and ``simplicial
gravity'' approaches \cite{3,4,5}.  While it could be argued that
both formulations involve Regge calculus and simplexes,
the distinguishing feature is that the former has a fixed
incidence matrix and varying link lengths while the latter
has a varying incidence matrix and fixed link lengths.

Both formulations are technically and computationally demanding.
For example, the Regge approach involves calculating areas and
deficit angles involving general d-simplexes.  In the
simplicial approach these take on only a limited number of possible
values, but the updating moves involve complicated interchanges of
several simplexes at once.  We introduce here a third lattice
gravity approach
which is structurally and computationally much simpler
than either Regge or simplicial
gravity, and may as a result be amenable to analytic attack
in more than two dimensions.

We call our formulation ``Ising-link quantum gravity''.  It is
easy to define.  The incidence matrix is fixed exactly as
in the Regge approach.  But the link lengths can only take
on two values,

$$ l_i = 1 + b s_i    \eqno(1) $$

with $s_i = \pm 1$ and $b$ a
positive constant.\footnote[4]{\hskip2pt $l_i = c(1 + b s_i)$
is no more general, as c can be absorbed into the definitions
of $\lambda$ and $k$ in (2).}  $i$ is a link label.  In order that
the triangle inequality (or its higher-dimensional
generalization - that the simplex volume is real and positive)
is always satisfied, it is straightforward
to show that we must take $b < {1 \over 3}$ in two dimensions,
$b < 3 - \sqrt{8} \approx .17$ in three dimensions, etc.
(See Section II.)  We restrict
b to satisfy this inequality so that {\it{all}} \hskip3pt
$2^{N_1}$ configurations are allowed.   ($N_1$ is the number of
links.)  This is
quite different from either Regge or simplicial gravity where most
potential updates either violate the triangle inequalities
or violate the manifold property.  Furthermore it provides us with a
natural measure which gives all $2^{N_1}$ configurations
equal weight.  It is clear that
our model is completely equivalent to a (regular lattice) Ising model
with spins ($s_i$) living on the links.  We will see that the spin
interactions are local, albeit somewhat complicated.

The Ising-link model is analytically and computationally much
simpler than either the Regge or simplicial gravity approaches.
But is it too simple?  In Section III we present mean field theory
results on the model in three-dimensions and
in Section IV we give corresponding Monte Carlo results.  We
compare to results obtained by Hamber and Williams for the
Regge theory in 3-d.

\vskip20pt
{\bf II. ISING MODEL FORMULATION}
\vskip15pt

In this section we discuss how to compute the Ising action
corresponding to the discrete form of

$$S = \lambda V - {k \over 2} \int d^dx \sqrt{g}\hskip2pt R
\hskip5pt ,
\eqno(2) $$

where V is the d-dimensional volume, $\int d^dx \sqrt{g}$
\hskip1pt, and
$R$ is the scalar curvature.  The lattice is formed out of hypercubes
plus face, cubic ($d \ge 3$) and hypercubic diagonals ($d \ge 4$),
with periodic boundary
conditions [2].  First we will treat two dimensions,
then three.  Four dimensions is just like three, only harder.

{\underbar{Two Dimensions}}: Consider a triangle with
link lengths $l_1$,
$l_2$ and $l_3$.  Define $l_i = 1 + b s_i$ as in (1).
Since $s_i^2 = 1$ and the formula
for the area of the triangle must be symmetric in the 3 spins,
the most general form for the area is

$$A_{123} = C_0 + C_1 (s_1 + s_2 + s_3) +
C_2 (s_1 s_2 + s_2 s_3 + s_3 s_1)
 + C_3 s_1 s_2 s_3 \hskip5pt . \eqno(3) $$

There are only 4 possible values for the area of the triangle,
corresponding to 0, 1, 2 or all 3 of its spins being equal to +1.
Computing these 4 areas and comparing to (3) gives four linear
equations
for the $C_\alpha$ in terms of the parameter b.  Their solution is

$$32 C_0 = 2 \sqrt{3} (1 + b^2) + 3 f(b) + 3 g(b) $$
$$32 C_1 = 4 b \sqrt{3} + f(b) - g(b) $$
$$32 C_2 = 2 \sqrt{3} (1 + b^2) - f(b) - g(b) \eqno(4) $$
$$32 C_3 = 4 b \sqrt{3} - 3f(b) + 3 g(b) \hskip5pt , $$

where $f(b) \equiv |1 - b| \sqrt{(1+3b)(3+b)}$ and
$g(b) \equiv (1 + b) \sqrt{(1-3b)(3-b)}$.  For example, $b = .1$
gives
$C_1 \approx .0291$, $C_2 \approx .0039$
and $C_3 \approx -.0008$.  As stated in the Introduction,
it is seen that we must have $b < 1/3$ for the
triangle areas to be real and positive.

Since the Einstein term in (2) is a topological invariant, it is not
relevant to the case of fixed topology being considered here.
Thus, summing over all triangles and dropping the irrelevant
constant term, the two-dimensional action is

$$S = \lambda (2 C_1 \sum_{i} s_i + C_2 \sum_{<ij>} s_i s_j
+ C_3 \sum_{<ijk>} s_i s_j s_k ) \hskip 5pt \hskip5pt, \eqno(5)$$

where $i$, $j$ and $k$ are link labels.
$<ij>$ indicates $i$ and $j$ are two of three links forming a
triangle.
In this case $s_i$ and $s_j$ may be termed nearest-neighbor links.
$<ijk>$ means that $i$, $j$ and $k$ are three links which form a
triangle.
The $C_2$ term is a ``nearest-neighbor'' ferromagnetic interaction.
The $C_1$ term is a magnetic field term and the $C_3$ term
is an additional symmetry-breaking term.

One might hope that the continuum limit of Ising-link quantum gravity
would correspond to a second-order magnetization phase transition.
But
with the explicit symmetry-breaking terms in (5), it is clear that
this transition can not be from order ($<s> \ne 0$) to disorder
($<s> = 0$).  It would have to be an $<s> \ne 0$ to $<s> \ne 0$
transition.
In two dimensions, as expected, we found no evidence of such a
transition, at least in the mean field theory approximation.

{\underbar{Three Dimensions}}: We will now go on to discuss the
form of the theory in three
dimensions.\footnote[5]{One reason for not dwelling on
the two-dimensional theory is that there is one (though only one)
known disagreement between 2-d Regge quantum gravity (which appears
most closely related to the 2-link model) and continuum results
\cite{6},
namely the critical exponents for Ising spins coupled to 2-d
Regge gravity \cite{7}.  One of the authors (M.G., unpublished) has
examined coupling Ising spins within a fixed invariant distance
rather
than those connected by a link, since the latter coupling
(used in [7])
has no invariant meaning.  But even with the invariant coupling
there
was still no agreement
between Regge gravity and continuum results; despite the fact
that the total area was held fixed, enough of
the links got sufficiently small for a finite fraction of the spins
to all become coupled together.  As a result the free energy
of that model is proportional to the number of degrees of freedom
{\it squared} - a  fatal illness.}
In the next section we will compare numerical results in 3-d
to results for the unconstrained Regge theory.

Consider the labeled tetrahedron of Fig. 1.  The volume $V_{tet}$ is
given by the formula [2]

\begin{eqnarray*}
\hskip10pt 144 V_{tet}^2 &=& 4 l_1^2 l_3^2 l_4^2 -
l_1^2 (l_3^2 - l_6^2 + l_4^2)^2
- l_3^2 (l_1^2 + l_4^2 - l_5^2)^2 - l_4^2 (l_1^2 - l_2^2 + l_3^2)^2 \\
&+& (l_3^2 - l_6^2 + l_4^2) (l_1^2 + l_4^2 - l_5^2)
(l_1^2 - l_2^2 + l_3^2)
\hskip2pt , \hskip174pt (6)
\end{eqnarray*}

where the $l_i$ may be written in terms of spins $s_i$ using Eq.
(1).  There are only 11 distinct possible values for the volume of the
tetrahedron, corresponding to the 11 a priori unknown constants in the
most general equation for $V_{tet}$ compatible with the symmetries of
the labeled tetrahedron:

\begin{eqnarray*}
\hskip50pt V_{tet} &=& C_0 + C_1 \sum_{i} s_i +
C_2 \sum_{<ij>} s_i s_j +
C_3 \sum_{[ij]} s_i s_j \\
&+& C_4 \sum_{<ijk>} s_i s_j s_k +
C_5 \sum_{(ijk)} s_i s_j s_k + C_6 \sum_{[ijk]} s_i s_j s_k
\\ &+& ( \prod_{l=1}^6 s_l) (C_0' + C_1' \sum_{i} s_i +
C_2' \sum_{<i,j>} s_i s_j + C_3' \sum_{[i,j]} s_i s_j )
 \hskip5pt . \hskip87pt (7)
\end{eqnarray*}

Here $<i,j>$ are again two of three links that form a triangle
in Fig. 1.  $[i,j]$ are the remaining pairs of links.
$<i,j,k>$ form a triangle, $(i,j,k)$ share a common site and
$[i,j,k]$ are the remaining triplets of links. Because $s_i^2 = 1$,
the last four terms involve 4,5 and 6-link interactons.
Evaluating
(7) for each of the 11 distinct volumes results in 11 linear
equations
for the $C_i$ and $C_i'$.  As in (4) for the 2-d case, these can
easily
be solved
to determine the $C_i$ and $C_i'$ as functions of $b$.  The
result is not particularly
illuminating and we will not reproduce it here.  It is worth noting,
however,
that the the volumes are always real and positive if we choose
$b < 3 - \sqrt{8} \approx .17$.  In the mean field and numerical
results
described in the next two sections, $b$ is held equal to $0.1$.

We see that after summing up the volumes of all the tetrahedrons,
the volume term in (2) will consist of only local interactions of
the spins,
involving up to 6-spin interactions.

The second (Einstein) term in (2),
$- {k \over 2} \int d^3x \sqrt{g}\hskip2pt R$,
takes the form [2]

$$S_E = k \sum_i l_i ( \sum_{t/i} \theta_{t/i} \hskip3pt - 2 \pi )
\eqno(8) $$

where $t/i$ denotes a tet containing the link $i$ and $\theta_{t/i}$
is the corresponding dihedral angle at link $i$.  For the tetrahedron
shown in Fig. 1, $\theta_{t/5}$ is given by

$$cos (\theta_{t/5}) =
{1 \over 16 A_{145} A_{256}} [ 2(l_4^2 + l_6^2 - l_3^2) l_5^2
- (l_4^2 + l_5^2 - l_1^2) (l_5^2 + l_6^2 - l_2^2) ] \hskip5pt ,
\eqno(9)$$

where $A_{ijk}$ is the triangle formed by links $i$, $j$ and $k$.
The
term $S_E$ can also be written in terms of local spin interactions,
but we
shall omit the details here.

\vskip20pt
{\bf III. MEAN FIELD THEORY IN THREE DIMENSIONS}
\vskip15pt

The Ising-link model is quite accessible to mean field theory (MFT)
techniques.
We write

$$ Z = \sum_{\{s\}} exp(-\beta H[s]) \hskip5pt , \eqno(10) $$

where here $\beta \equiv 1$ and

$$ H \equiv S = \lambda V -
{k \over 2} \int d^3x \sqrt{g}\hskip2pt R \hskip5pt .
\eqno(11) $$

This is a functional of the spins since $l_i = 1 + b s_i$, by Eq. (1).
We wish to minimize the free energy,

$$ F = \hskip4pt < H > - \hskip4pt \cal{S}/\beta \hskip5pt ,
\eqno(12) $$

for the spin probability distribution function, $P[s]$, where
$\cal{S}$
is the entropy.

The mean field approximation [8] consists of replacing the true
probability
distribution for the spins by a factorized form:

$$ P[s] \rightarrow p(s_1) p(s_2) p(s_3) ... p(s_{N_1}) \hskip5pt ,
\eqno(13) $$

where $N_1$ is the number of links.
If all links were equivalent,
we could write $p(s_i) = { 1 + m s_i \over 2} \Rightarrow
\sum_{s_i} p(s_i) = 1$ and
$<s_i> \hskip3pt  = \hskip2pt m$.  But there are three different
kinds of links
in the lattice formed out of cubes with body and face diagonals:
the body diagonals, the cube edges, and the
face diagonals.  Links of the same type have the same
geometrical environment; links of different types don't.  As a
result
we must use the more general distribution,

$$ p_j(s_i) \equiv {1 + m_j s_i \over 2} \hskip5pt, \eqno(14) $$

where $j = 1,2$ and $3$ for body diagonals, cube edges and face
diagonals
respectively.

A straightforward calculation allows us to determine $<H>$ and
$\cal{S}$
as functions of $m_1$, $m_2$ and $m_3$.  Let

$$ P_V \equiv p_2(s_1) p_3(s_2) p_2(s_3) p_3(s_4) p_1(s_5) p_2(s_6) $$
$$ P_{R1} \equiv
p_1(s_1) p_3(s_2) p_2(s_3) p_3(s_4) p_2(s_5) p_2(s_6) $$
$$ P_{R2} \equiv
p_3(s_1) p_2(s_2) p_1(s_3) p_2(s_4) p_2(s_5) p_3(s_6)
\eqno(15)$$
$$ P_{R3} \equiv p_2(s_1) p_1(s_2) p_3(s_3) p_2(s_4) p_3(s_5) p_2(s_6)
\hskip5pt .$$

We find that
$$<V> = 6 N_0 \sum_{s_1} ... \sum_{s_6} P_V(s) V_{tet}(s)
\hskip5pt, \eqno(16)$$
\begin{eqnarray*}
<{1 \over 2} \int d^3x \sqrt{g}\hskip2pt R> &=&
2 \pi N_0 [ 7 + b(m_1 + 3 m_2 + 3 m_3) ] \\ -
&6 N_0& \hskip-10pt \sum_{s_1} ... \sum_{s_6} l_5(s) \theta_{t/5}(s)
[ P_V(s) + 2 P_{R1}(s) + P_{R2}(s) + 2 P_{R3}(s) ] \hskip5pt
\hskip19pt (17)
\end{eqnarray*}
and
$$ {\cal{S}} \equiv - <ln P[s]> \hskip5pt =
-N_0 [ h(m_1) + 3 h(m_2) + 3 h(m_3) ] \hskip5pt , \eqno(18) $$

where $V_{tet}$ and $\theta_{t/5}$ are given by (6) and (9)
respectively,
$N_0$ is the number of lattice sites and
$h(x) \equiv { 1 + x \over 2} \hskip3pt ln( { 1 + x \over 2} ) +
{ 1 - x \over 2} \hskip3pt ln( { 1 - x \over 2} ) $.  $s$ is
shorthand
for $s_1,s_2,...,s_6$.
Now it is a simple matter to numerically minimize the free energy
(12)
as a function of $m_1$, $m_2$ and $m_3$.  Then $<s>$ is given by

$$ <s> = {m_1 + 3 m_2 + 3 m_3 \over 7} \hskip5pt , \eqno(19)  $$

and

$$ {\cal{R}} \equiv
<l^2> {<\int d^3x \sqrt{g}\hskip2pt R> \over <V> } =
(1 + b^2 + 2 b <s>) { <\int d^3x \sqrt{g}\hskip2pt R> \over <V> }
\hskip5pt . \eqno (20) $$

Here we follow the notation of Hamber and Williams [9].  Also the
``curvature susceptibilty'' is defined as

$$ \chi_R =
{2 \over < V >} {\partial \over \partial k}
<\int d^3x \sqrt{g}\hskip2pt R>
\hskip5pt . \eqno(21) $$

{\underbar{Results}}. \hskip5pt Fig. 2 shows typical MFT results
for the case
$\lambda = 1$.  There is sharp cross-over behavior seen in
$<s>$, $v$
and ${\cal{R}}$ at $k$ slightly negative.  ${\cal{R}}$ and
$<\int d^3x \sqrt{g}\hskip2pt R>$ (not shown in Fig. 2) are
related by (20)
and exhibit similar behavior in this region.
$\chi_R$ was evaluated using (21), by taking a numerical
derivative of $<\int d^3x \sqrt{g}\hskip2pt R>$
with increment $\Delta k = 1$.
We see a peak in $\chi_R$
here, as expected from the rapid cross over behavior in ${\cal{R}}$.
Hamber and Williams [9] found a second order phase transition
in the 3-d Regge theory exhibiting

$$ \chi_R \sim |k_c - k|^{\delta -1} \eqno(22)$$

for $k < k_c$, with $\delta = 0.80 \pm 0.06$,
a very weak second order phase transtion.
To investigate whether this kind of non-analyticity is
seen in the MFT approximation to the Ising-link model, we varied
$\Delta k$ from $1$ downward.  ($\Delta k$ is the increment used
to take the
numerical derivative of $<\int d^3x \sqrt{g}\hskip2pt R>$.)
The behavior (22) would result in the peak of $\chi_R$ growing with
$\Delta k$ like $(\Delta k)^{\delta - 1}$.  $\delta = 0.80$ would
imply that
the peak would grow by $58\%$ in height for each factor of $10$
decrease in $\Delta k$.  However, for all values of $\lambda$
considered
(up through $\lambda = 75$),
there was {\it no} increase in the peak height
as $\Delta k$ was decreased from $1$ down to $.01$.  As a result
we have no evidence of (22) with $\delta - 1 < 0$
in the {\it mean field approximation} to the 3-d Ising-link model.
Nevertheless, in the next
section we will present evidence {\it Monte Carlo} evidence
for (22) with $\delta - 1 < 0$ at large values of $\lambda$.

For $\lambda \ne 1$ the situation looks similar to
that shown in Fig. 2.  There is a finite peak in $\chi_R$
at $k$ slightly
negative and a first order phase transition at large positive $k$.
The height of the peak in $\chi_R$ and the size of the discontinuity
at the first order phase transition vary appreciably with $\lambda$.
Fig. 3 shows a dashed curve in $\lambda$ - $k$ space where there is
a peak in $\chi_R$ and a solid curve of first order phase transitions.
For $k$ to the right of the first order phase transition, ${\cal{R}}$
rapidly approaches
zero from below as the system approaches a state with
$m_1 = m_3 = - m_2 = 1$.
As $k \rightarrow -\infty$, the system approaches a state with
$m_1 = m_2 = - m_3 = 1$.

\vskip20pt
{\bf IV. NUMERICAL RESULTS IN THREE DIMENSIONS}
\vskip15pt

The 3-d Ising-link model was also analyzed by the Monte Carlo method.
The discrete form of (2) was used, with toroidal topology
and without higher derivative terms, as discussed in the previous
section.
As in that section, $b$ was chosen to be $0.1$ thoughtout.
(See Eq. (1).)

Since there are only two possible lengths per link, there are at
most $2^6 = 64$ possible configurations for a particular tetrahedron.
(Actually there are significantly fewer due to symmetry.)
Because all terms in the action are determined solely by the
link lengths in the lattice, the limited number of distinct
tetrahedrons allows many of the calculations to be performed
only once at program entry and stored for later use in the form
of ``look-up tables''.  These tables are accessed during the Monte
Carlo updating.
As a result, the Ising-link model proved to be
quite computationally efficient; run times were reduced by as
much as a factor of ten over continuous-link (Regge gravity)
simulations.

Except for the reliance on look-up tables, the simulations were
carried
out in the usual way.  An initial random configuration of link
lengths is chosen,
generating a particular initial geometry.  A link update consists of
choosing a particular link
in the lattice, calculating the change in the action if the link
takes on its other possible value, and accepting the new link
value with probability proportional to the
exponential of the negative change in the action (heat bath).  Link
updates are performed for each link in the lattice; this constitutes
one
sweep.  The quantities of interest are calculated after each
sweep of the lattice, and the values for each new geometry are
binned for statistical analysis.  Runs of up to 100k sweeps on
the $4^3$ and $8^3$ lattices, and 80k sweeps on the $16^3$
lattice were performed for various values of $\lambda$ and $k$.

The two physical quantities of greatest interest were ${\cal{R}}$
and
$\chi_R$, defined in Eqs. (20) and (21).
Hamber and Williams found that in the Regge theory, the
curvature susceptibility diverges at points where ${\cal{R}}$
vanishes [9].
Thus, a portion of the curve ${\cal{R}} = 0$ was first
identified (Fig.
4), and the behavior of the model was studied along that curve.
But peaks in
$\chi_R$ did not appear along the ${\cal{R}} = 0$ curve, but rather
were
located at values of $k$ that correspond to local inflection
points in ${\cal{R}}$.  This behavior is consistent with Eq. (21),
which relates $\chi_R$ to the first derivative of ${\cal{R}}$ with
respect to $k$.  Also plotted in Fig. 4 is a dashed curve of
peaks in $\chi_R$.  The case $\lambda = 1$ was
studied for values of $k$ close to that curve, but no growth of
the  peak with increasing system size was observed, indicating
the absence of a second-order transition in this region of
parameter space.  At $\lambda = 75$, however, we did find growth in
the
$\chi_R$ peak with increasing system size indicative of a second
order phase transition (see below).  We expect
that the dashed curve becomes a line of second order phase
transitions somewhere
between $\lambda = 1$ and $\lambda = 75$.

Fig. 4 may be compared with Fig. 3 determined by MFT.  The location
of the
peak in $\chi_R$ is the same in both plots to within Monte Carlo
statistical
errors.  However no
second order phase transition occured in the MFT approximation for
any
value of $\lambda$.  The agreement for large $k$ between MFT and
Monte Carlo
is poorer.  As discussed
in the previous section, MFT exhibited a first order phase
transition along the
solid curve of Fig. 3, and for $k$ to the right of that curve,
${\cal{R}}$ asymptotically approached $0$ from below.
But no first order phase transition was found in
Monte Carlo, and  ${\cal{R}}$ went from negative to positive
values at the
location of the solid curve in Fig. 4.

The comparison with MFT is also seen by comparing Monte Carlo
$\lambda = 1$ data displayed in Fig. 5 with corresponding MFT
data shown
in Fig. 2.  Note the difference in the scales for $< s >$ and
$\chi_R$.
For $k$ negative the agreement is quite good; in fact the Monte Carlo
results agree completely with MFT at $k \rightarrow -\infty$;
both indicate that the system freezes into a state with the body
diagonals
and cube edges long ($s = 1$) and the face diagonals short ($s = -1$).
The main negative $k$ disagreement occurs at the peak in $\chi_R$
which is much
lower in the MFT approximation than even on a $4^3$ lattice.  For $k$
positive the agreement is much poorer.  For large $k$ in MFT,
${\cal{R}}$ never goes positive,
a spurious first order phase transition is predicted
($k \approx 48 $)
and $< s >$ is off by about a factor of 3.

We now return to the evidence for a second order phase transition at
large $\lambda$.  The desired signature for critical behavior would
be

$$ {\cal{R}} \approx {\cal{R}}_0 + A |k_c - k|^{\delta} \eqno(23)$$

and

$$ \chi_R \approx B |k_c - k|^{\delta -1} \eqno(24)$$

for $k \approx k_c$,
where $k_c$ is the critical point for fixed $\lambda$ and $\delta$
is the critical exponent characteristic of the transition [9].

Previous work on the full Regge theory in three dimensions at
$\lambda = 1$
determined that ${\cal{R}}_0 \approx 0$ and
$\delta = .80 \pm .06$ [9].  In that theory $k$ had to
approach $k_c$ from below; the theory was sick for $k > k_c$.
Here we found that the Ising-link model
has ${\cal{R}}_0 \ne 0$ at the peaks in
$\chi_R$.
The curvature susceptibility, though, does
show behavior expected of a second-order phase transition.  Near
criticality there is an observed narrowing in the curvature
susceptibility and an increase in peak height
with increasing system size.  This is readily seen in Fig. 6.
Following Hamber and Williams [9], the finite-size scaling relation
for
the peak of the
curvature susceptibility is

 $$ ln(\chi_R) \sim c + {\alpha \over \nu} \hskip2pt ln L
\hskip5pt ,
\eqno(25)$$

where $L$ is the system length and
$\alpha / \nu = d (1 - \delta)/(1 + \delta)$, with $d = 3$.
Using this relation, the critical exponent was determined from
the curvature susceptibility data (Fig. 6) to be
$\delta =.55 \pm .02$
for the Ising-link model with $\lambda = 75$.

We conclude that at least at this one value of $\lambda$, the
Ising-link
model does appear to undergo a second order phase transition of the
form considered by Hamber and Williams.  However
the two values of $\delta$ are statistically incompatible, indicating
two distinct universality classes for the two models.

\vskip20pt
{\bf V. DISCUSSION AND SUGGESTIONS FOR FURTHER WORK}
\vskip15pt

In three dimensions we have uncovered critical behavior in the
Ising-link model for $\lambda = 75$ but not for $\lambda = 1$.
The region
between should be carefully investigated.  At $\lambda = 75$ the
critical
behavior takes the form (23) and (24) as found by Hamber and
Williams
for the full Regge theory in 3-d.  But ${\cal{R}}_0$ was $0$ in their
model and not in ours.  Their model was sick for $k > k_c$; ours
was not.  At $\lambda = 75$ our critical exponent $\delta$ is
statistically
different from that seen by Hamber and Williams at $\lambda = 1$
(where we
found no phase transition).
Does this difference persist at other values of
$\lambda$ or do both models exhibit universality in $\lambda$?
Is there
universality in $b$ for the Ising-link model?
Clearly more work in three dimensions is needed.

The Ising-link model can and should be investigated in four
dimensions as well.
The mean field approximation is somewhat more difficult but still
quite feasible,
and it should be more accurate in four dimensions than in three.
Other analytic methods should also be considered.
4-d Monte Carlo computations can still be performed using look-up
tables and
as a result should be many times faster than for the full Regge
theory.
If different universal behavior between the two theories persists
in
four dimensions, more work would be needed to determine which,
if either theory
is relevant to the universe we live in.

\vskip20pt
{\bf ACKNOWLEDGEMENTS}
\vskip15pt

M.G. is very grateful to H. Hamber for many helpful discussions.
This work
was supported in part by the National Science Foundation
through Grant No. NSF-PHY-9007497, and California State University.

\vskip20pt
{\bf Note added}:
While writing up this work we received two related preprints by
W. Bierl,
H. Markum, and J. Riedler [10].  They independently define
the Ising-link model and investigate its behavior in two dimensions.

\bibliographystyle{unsrt}

\vskip50pt
{\Large\bf Figure Captions}
\vskip3pt

Fig. 1.  A labeled tetrahedron.

Fig. 2.  A plot of the indicated quantities versus $k$ for
$\lambda = 1$
as calculated in MFT.  \hskip2pt $v$ is the average volume per
site, $<V>/N_0$.

Fig. 3.  The `phase diagram' of the 3-d Ising-link model in the
MFT approximation.
The dashed curve shows the location of the peak in $\chi_R$ and
the solid
curve is a first order phase transition.

Fig. 4.  The `phase diagram' of the 3-d Ising-link model as
determined by Monte
Carlo simulations.  The dashed curve shows the location of the
peak in $\chi_R$
which is found to scale with system size at large $\lambda$.
The solid curve is
${\cal{R}} = 0$.  Lattice sizes up to $16^3$ were used, and error
bars are of order
the size of the data points.

Fig. 5.  Monte Carlo data for $\lambda = 1$ on a $4^3$ lattice.
Data was
taken every $\Delta k = 1$, and error bars are 0.013 or smaller.

Fig. 6.  The peak in $\chi_R$ for lattices of length $4$, $8$ and
$16$.
The largest statistical errors of the data points are respectively,
0.3, 1.3 and 3.3.

\end{document}